# Evolution of Effective Temperature, Kinetic Freeze-out Temperature and transverse flow velocity in pp Collision

Murad Badshah[1], Muhammad Ajaz[1,*], Muhammad Waqas[2], Hannan Younis[3]

[1]Department of Physics, Abdul Wali Khan University Mardan, 23200 Mardan, Pakistan
[2]School of Mathematics, Physics and Optoelectronic Engineering, Hubei University of Automotive Technology, Shiyan 442002,
China
[3]Department of Physics, COMSATS University Islamabad, 44000 Islamabad, Pakistan
*Corresponding author: muhammad.ajaz@cern.ch; ajaz@awkum.edu.pk (M. Ajaz)

## Abstract

This article focuses on the study of strange hadrons ($K_S^0$, $\Lambda$, $\overline{\Lambda}$, $\Xi^+$, $\Xi^-$) at $\sqrt{s} = 0.2$ TeV, recorded by STAR at RHIC, and at $\sqrt{s} = 0.9$ TeV, 5.02 TeV and 7 TeV, recorded by CMS at LHC, in pp collision in the rapidity range from 0 to 2. The $p_T$ distributions of these strange particles have been processed using two statistical models, the Tsallis and the modified Hagedorn model. Both models fit the experimental data well. We extracted different freezeout parameters from the fit procedure using the abovementioned functions. We found that with increasing the collision energy, the effective temperature (T), in the case of the Tsallis model, and kinetic/thermal freeze-out temperature ($T_0$) and transverse flow velocity ($\beta_T$), in the case of the modified Hagedorn model, increase because of greater energy transfer among the participants at higher colliding energies. Both T and $T_0$ are observed to increase with the increase in the rest masses of the outgoing particles revealing the multi-freeze-out scenario. Furthermore, the multiplicity parameter ($N_0$) decreases with the increase in the particle mass, confirming the mass differential freeze-out scenario. An inverse relationship between the non-extensivity parameter (q) and the masses of the produced particles has been noticed. Similarly, an inverse correlation between q and T has been found. For lighter particles, smaller T and greater q mean that they decouple from the system later and attain equilibrium slowly compared to heavier ones. In addition, a positive correlation between $\beta_T$ and $T_0$ is noticed, which agrees with the literature.

## Introduction

The collisions of heavy nuclei at high energies produce the deconfined state of matter known as Quark-Gluon Plasma (QGP), where quarks and gluons are assumed free to interact. The direct detection of QGP has yet to be made possible due to its short lifetime. However, there are many signatures for its existence like strangeness enhancement, high thermal effective temperature, hardening of transverse momentum ($p_T$) spectra, etc which are all observed in heavy ions or nuclei collisions like in Pb-Pb and Au-Au collisions leading to the discovery of QGP. Once the QGP is formed, to gain thermal equilibrium with its surroundings, it expands and cools down, during expansion and cooling there are two main freeze-out stages. The stage at which inelastic collision between particles of the system ceases, elastic interactions start, and the ratio of the particles turns to be fixed is called the chemical freeze-out stage, and the corresponding temperature of the system is called chemical freeze-out temperature. After the chemical freezeout stage, the stage of thermal or kinetic freeze-out arrives, where the corresponding temperature is called thermal or kinetic freeze-out temperature. In this stage, the mean free path of these particles becomes greater than the size of the system, they get separated from the parent system and move towards the detectors.

Some of the signatures of QGP have also been observed in high multiplicity p-p collisions at the Large Hadron Collider (LHC) and Relativistic Heavy Ion Collider (RHIC). These include strangeness production [1], increased yield of particles [2], equivalent effective temperature (T), kinetic freeze-out temperature ($T_0$) and transverse flow velocity ($\beta_T$) to those obtained in nucleus-nucleus (A-A) collisions [3, 4] and multiparticle ridge-like correlations [5]. The p-p collisions provide bases for studying A-A and p-A collisions. In the present work, we analyzed the $p_T$ distribution of strange hadrons. The $p_T$ spectra give information about the system during its freeze-out [6]. Moreover, searching for the strange hadron production in pp collision is very significant as strangeness enhancement is one of the signatures of QGP.

The non-extensive Tsallis distribution function, which is a simple function with just two parameters, namely T and the non-extensive parameter (q), and can cover the wide $p_T$ range, can be used to fit the experimental data for $p_T$ distribution. In fact, the effective temperature is the combination of kinetic freeze-out temperature and radial flow where the former gives information about the equilibration of the system, while the latter gives information about the expansion of the system [7]. We can extract both parameters from the Tsallis distribution function by an alternative method. Still, the scale used in extracting the effective temperature, kinetic freeze-out temperature, and radial flow are not the same. Different methods have different scales to extract the parameters. Therefore we choose the Hagedorn model with the embedded flow to extract the parameters directly from the transverse momentum spectra of the particles. We also have other choices to choose from other models (such as the Blast wave model with Boltzmann Gibbs statistics [8-10], Blast wave model with Tsallis statistics [11, 12], Tsallis distribution with transverse flow effect [13, 14] and modified Hagedorn function with embedded $\beta_T$ [15-17]) for the extraction of kinetic freeze-out temperature and radial flow but they can not be used for the wide $p_T$ range. For instance, the Blast wave model with Boltzmann Gibb's statistics can be used for the low $p_T$ of 0-2 or 2.5 GeV/c, and the Blast wave model with Tsallis can be used up to a $p_T$ range of 0-6 GeV/c.

## Experimental Data and Models

The experimental data describing the $p_T$ distribution in the rapidity range, 0 < y < 2 for pp collision at $\sqrt{s} = 0.2$ TeV, recorded by STAR at RHIC, and at $\sqrt{s} = 0.9$ TeV, 5.02 TeV and 7 TeV, recorded by CMS collaboration at LHC, is taken from [18], [20], [19] and [20] respectively. The experimental data of the aforementioned references have been downloaded from the website hepdata.net.

The Tsallis distribution function (model) used to describe $p_T$ spectra of particles produced in high energy collisions in its simplest form is given in Eq. 1.

$$\frac{d^2N}{N_{ev}\, dp_T\, dy} = 2\pi C\, p_T \left(1 + (q-1)\frac{m_T}{T}\right)^{-1/(q-1)} \tag{1}$$

Where $C$ is the normalization or fitting constant which is assumed to be linearly related to the size or volume of the system i.e., $C = gV/(2\pi)^3$ where $g$ represents the degeneracy factor and $V$ is the volume of the system [7], $q$ is the free parameter called non-extensivity parameter and $m_T$ is the transverse mass, given by, $m_T = \sqrt{p_T^2 - m_0^2}$, where $m_0$ is the rest mass of the particle under consideration.

Among many other forms of Tsallis function the thermodynamically most consistent Tsallis function, which has also been used in our coding in the present work, is given in Eq. 2. [21-23].

$$\frac{d^2N}{N_{ev}\, dp_T\, dy} = 2\pi C\, p_T\, m_T \left(1 + (q-1)\frac{m_T}{T}\right)^{-q/(q-1)} \tag{2}$$

To describe the $p_T$ distribution of particles in high-energy collisions the Hagedorn function (model) is also very efficient, Eq. 3 represents one of the simplest forms of this function [24].

$$\frac{d^2N}{N_{ev}\, dp_T\, dy} = 2\pi N_0\, p_T \left(1 + \frac{m_T}{p_0}\right)^{-n} \tag{3}$$

Where $N_0$ is the normalization constant and n and $p_0$ are the free parameters. In Eq. 1 and Eq. 2, T represents the effective temperature which includes contributions from kinetic freeze-out temperature and transverse flow velocity. To include these parameters in the Hagedorn function given in Eq.3 compare Eq. 1 and Eq. 3, one can easily deduce that the two equations are mathematically equivalent if, $n = 1/(q-1)$ and $p_0 = nT$, doing these substitutions Eq. 3 becomes,

$$\frac{d^2N}{N_{ev}\, dp_T\, dy} = 2\pi N_0\, p_T \left(1 + \frac{m_T}{nT_0}\right)^{-n} \tag{4}$$

Where $T$ has been replaced by $T_0$ to differentiate $T_0$ (kinetic freeze-out temperature) from $T$, the effective one. To include $\beta_T$, we use the simple Lorentz transformation, $m_T \sim <\gamma_T> (m_T - p_T <\beta_T>)$ which has already been used in the Ref. [15, 17]. Using this transformation for $m_T$ in Eq. 4, we get what is given in Eq. 5.

$$\frac{d^2N}{N_{ev}\, dp_T\, dy} = 2\pi N_0\, p_T \left(1 + \frac{<\gamma_T>(m_T - p_T<\beta_T>)}{nT_0}\right)^{-n} \tag{5}$$

Where $<\gamma_T> = \frac{1}{\sqrt{1 - <\beta_T>^2}}$ and $<\beta_T>$ is the mean transverse flow velocity of the expanding system. Eq. 5 is the modified Hagedorn function with the embedded $<\beta_T>$ and $T_0$ and has been used in our coding in the present work.

In our analysis, we employed the minimum $\chi^2$ method to fit the experimental data using Eq. 2 and Eq. 5. This method aims to find parameter values that minimize the $\chi^2$ value, thereby indicating the best agreement between the model's predictions and the experimental observations. For the purpose of this fitting, we considered only statistical uncertainties, which are uncorrelated. We opted not to include systematic uncertainties in the fit, as they are typically correlated across bins and their inclusion would require a more complex treatment to be fully accurate.

## Results and Discussion

Plots of Fig. 1 are obtained by fitting the Tsallis model with experimental data for $\sqrt{s} = 0.2 \, TeV, 0.9 \, TeV, 5.02 \, TeV$ and $7 \, TeV$ in pp collision. The solid lines represent the model fitting while the different colored markers are used for experimental data points. These plots represent the $p_T$ spectra of different strange particles, $K_S^0, \Lambda, \overline{\Lambda}, \Xi^+,$ and $\Xi^-$. The $p_T$ spectra along the Y-axis of all plots are scaled i.e., multiplied with some numbers, written on the top right corner of each plot, to enhance the distribution visibility. The plots show that the Tsallis model agrees well with the experimental data.

Plot (a) of Fig.1 represents the $p_T$ distribution of $K_S^0, \Lambda, \overline{\Lambda}, \Xi^+, \Xi^-$ at $\sqrt{s} = 0.2 \, TeV$. Plots (b) and (d) of the same figure represent the $p_T$ distribution of $K_S^0$, $\Lambda$ and $\Xi^-$ at $\sqrt{s} = 0.9 \, TeV$ and $\sqrt{s} = 7 \, TeV$ while plot (c) shows the $p_T$ spectra of $K_S^0$, $\Lambda + \overline{\Lambda}$ and $\Xi^+ + \Xi^-$ at $\sqrt{s} = 5.02 \, TeV$. The bottom panel of each plot shows data by fit ratio (Data/Fit), which measures the deviation of the fit line (model) from experimental data.

While fitting the Tsallis distribution function with the experimental data, we obtain the values of very important parameters needed to explain the freeze-out stage of the produced system like effective temperature (T), non-extensivity parameter (q), and fitting constant ($N_0$). The values of all these parameters along with the $\chi^2$ (deviation from experimental value) and NDF (number of degrees of freedom) are listed in Table 1. We tried to keep the $\chi^2$ as small as possible in all the fittings.

Fig. 2 (a) shows the relation of T with the masses of the observed particles and center of mass energy ($\sqrt{s}$), where T has been found to increase with increase in $\sqrt{s}$, because of the violent collisions at higher energies T of the produced particle increases, the earlier also increases with an increase in the masses of the outgoing particles. Which means that the heavier particles have greater T compared to lighter ones, which renders that the heavier particles will decouple from the system quickly and then followed by the lighter particles. This scenario is called multi-freeze-out scenario. Fig. 2 (b) shows the inverse correlation between q and masses of the produced particles ($m_0$) and plot (c) of the same figure shows the inverse relation of T and q, which means that lighter particles will decouple from the fireball later. Hance will attain equilibrium slowly because they have smaller T and larger q. These results are consistent with the results of our recently published paper [25]. Fig. 2 (d) is the plot between $N_0$ and masses of the produced particles, which shows that with increasing mass, $N_0$ or volume of the produced system (because the volume of the system is directly related to $N_0$) also increases, which again confirms that heavier particles decouple from the fireball quickly than the lighter particles and proves the mass differential freeze-out or multi-freeze-out scenario.

Plots (a), (b), (c), and (d) of Fig. 3 show the $p_T$ spectra of strange hadrons at $\sqrt{s} = 0.2 \, TeV, 0.9 \, TeV, 5.02 \, TeV$ and $7 \, TeV$ obtained through modified Hagedorn model with embedded $\beta_T$. As usual, different shapes represent data points and solid lines are used for the corresponding model. The result of the fit procedure on the experimental data points show that the Hagedorn model reproduce the experimental data. The fitting is done for the smallest possible value of $\chi^2$ and the values of various important parameters like kinetic freeze-out temperature ($T_0$), $\beta_T$, n, and fitting constant ($N_0$) are obtained. These values are listed in Table 2.

Plot (a) of Fig. 4 shows the relation between $T_0$ and masses of the observed strange particle. This plot also shows the relation between $T_0$ and $\sqrt{s}$. $T_0$ is found to increase with increase in mass and / or $\sqrt{s}$ because of the same reasons discussed above in case of T. Plot (b) of the same figure shows the relation of $\beta_T$ with $\sqrt{s}$, where $\beta_T$ is found to vary directly with $\sqrt{s}$. Because of the greater energy transfer in violent collision, the produced particles will have greater kinetic energy which results in the greater $\beta_T$. Fig. 4 (c) is used to show how $T_0$ varies with $\beta_T$, where the former is found to increase with increase in the later representing greater $\beta_T$ for those particles having greater $T_0$. Plot (d) of the same figure represents $N_0$ as a function of the masses of the produced strange particles having the same explanation as in the above paragraphs in case of Tsallis model.

| Energy [TeV] | Particle | T [MeV] | q | $N_0$ | $\chi^2$ | NDF |
|---|---|---|---|---|---|---|
| 0.2 | $K_S^0$ | 96.91±0.01 | 1.108±0.002 | 6.671±0.005 | 28.3595 | 20 |
| | $\Lambda$ | 98.44±0.03 | 1.079±0.001 | 1.532±0.001 | 13.2884 | 19 |
| | $\overline{\Lambda}$ | 98.44±0.03 | 1.079±0.001 | 1.399±0.001 | 19.7096 | 19 |
| | $\Xi^+$ | 125.98±0.04 | 1.072±0.001 | 0.078±0.0007 | 5.5589 | 9 |
| | $\Xi^-$ | 125.98±0.02 | 1.072±0.001 | 0.078±0.0004 | 4.6215 | 9 |
| 0.9 | $K_S^0$ | 104.52±0.02 | 1.125±0.003 | 12.351±0.009 | 19.1243 | 22 |
| | $\Lambda$ | 105.02±0.01 | 1.096±0.001 | 6.186±0.005 | 20.5432 | 22 |
| | $\Xi^-$ | 135.31±0.04 | 1.084±0.001 | 0.682±0.001 | 14.9712 | 20 |
| 5.02 | $K_S^0$ | 107.11±0.03 | 1.148±0.001 | 8.341±0.007 | 7.7407 | 30 |
| | $\Lambda + \overline{\Lambda}$ | 110.23±0.02 | 1.124±0.001 | 3.011±0.002 | 0.8943 | 14 |
| | $\Xi^+ + \Xi^-$ | 145.24±0.03 | 1.115±0.001 | 0.258±0.001 | 3.0185 | 8 |
| 7 | $K_S^0$ | 117.15±0.02 | 1.146±0.003 | 20.889±0.010 | 15.5437 | 22 |
| | $\Lambda$ | 135.24±0.03 | 1.112±0.002 | 11.201±0.004 | 18.6549 | 22 |
| | $\Xi^-$ | 155.34±0.02 | 1.108±0.001 | 1.289±0.001 | 16.7692 | 20 |

***Table 1.*** *The values of different free parameters, normalization constant, and $\chi^2$ for different strange hadrons at different collision energies were obtained from the Tsallis model.*

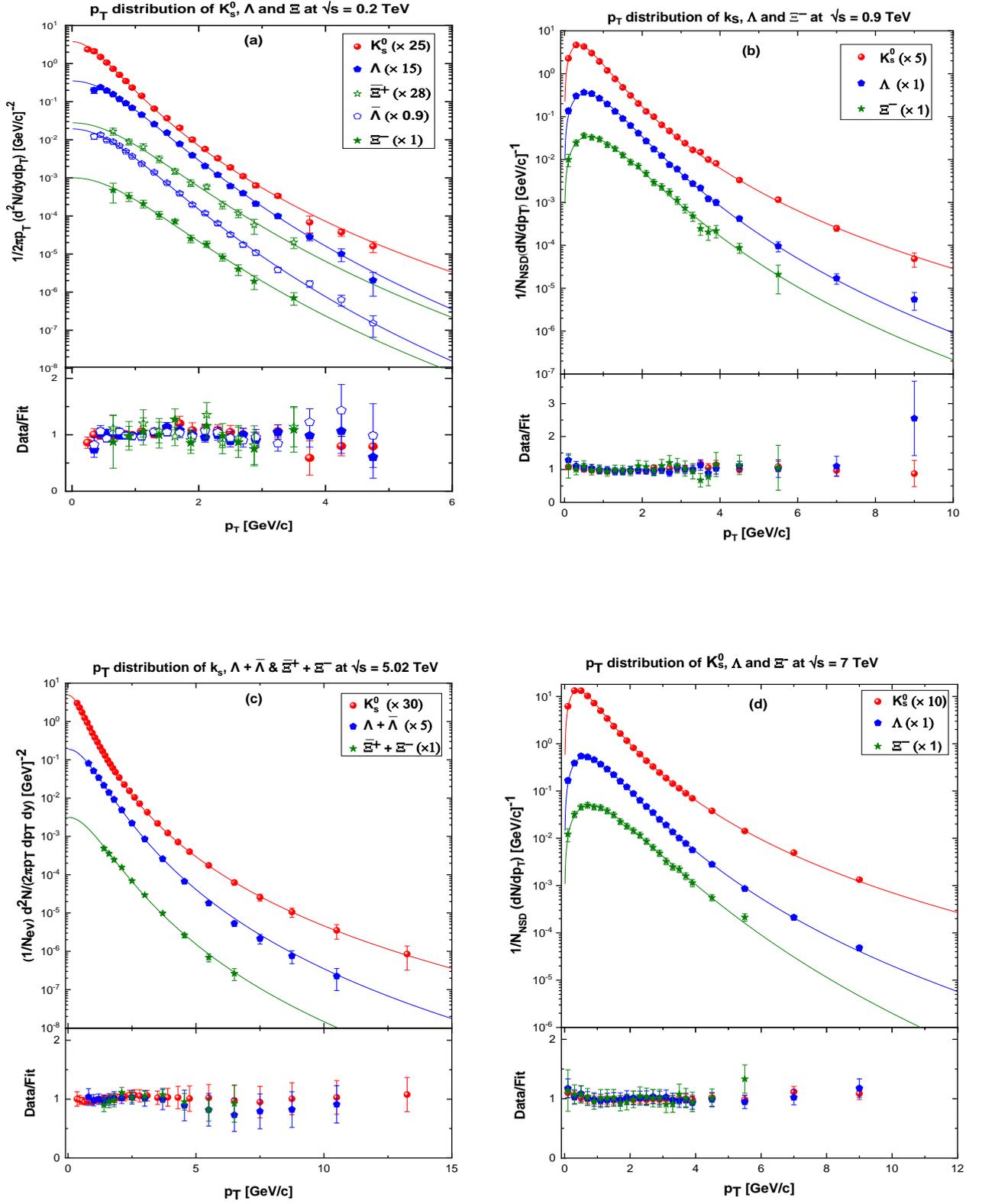

*Fig. 1.* $p_T$ *distribution of strange Hadrons at the different center of mass energies in pp collision, the symbols are used for data points and solid lines represent the fits of Tsallis model, the lower panel associated with each plot shows the data by fit ratio.*

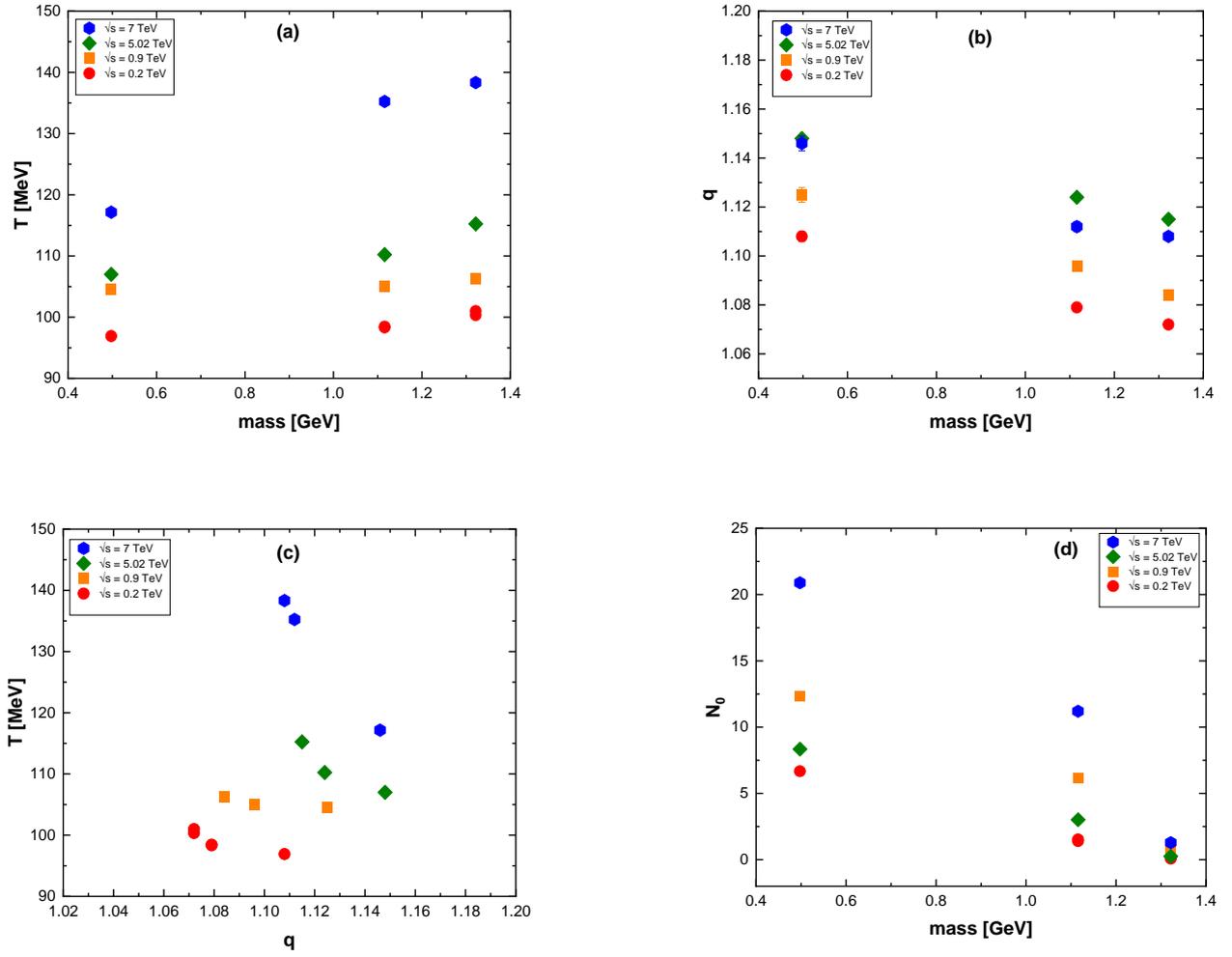

***Fig.2*** *Dependence of (a) T on rest masses of the observed particles ($m_0$), (b) q on $m_0$ (c) T on q and (d) $N_o$ on $m_0$ at different collision energies using Tsallis model.*

| Energy [TeV] | Particle | T [MeV] | $\beta_T$ [c] | n | $N_0$ | $\chi^2$ | NDF |
|---|---|---|---|---|---|---|---|
| 0.2 | $K_S^0$ | 67.96±0.01 | 0.19±0.01 | 8.17±0.01 | 6.631±0.001 | 59.1360 | 19 |
| | $\Lambda$ | 68.11±0.01 | 0.126±0.01 | 11.31±0.01 | 1.392±0.001 | 12.7337 | 18 |
| | $\overline{\Lambda}$ | 68.11±0.01 | 0.165±0.01 | 12.11±0.01 | 1.190±0.001 | 15.4776 | 18 |
| | $\Xi^+$ | 74.71±0.01 | 0.128±0.01 | 10.78±0.01 | 0.069±0.001 | 5.2491 | 8 |
| | $\Xi^-$ | 74.77±0.01 | 0.129±0.01 | 10.75±0.01 | 0.071±0.001 | 4.0610 | 8 |
| 0.9 | $K_S^0$ | 75.54±0.01 | 0.194±0.01 | 7.18±0.01 | 79.469±0.001 | 21.7656 | 21 |
| | $\Lambda$ | 76.78±0.01 | 0.133±0.01 | 9.71±0.01 | 38.301±0.001 | 17.7654 | 21 |
| | $\Xi^-$ | 77.73±0.01 | 0.162±0.01 | 10.04±0.01 | 4.201±0.001 | 18.7643 | 19 |
| 5.02 | $K_S^0$ | 80.01±0.01 | 0.201±0.01 | 6.07±0.01 | 8.001±0.001 | 14.7772 | 29 |
| | $\Lambda + \overline{\Lambda}$ | 82.48±0.01 | 0.166±0.01 | 7.62±0.01 | 2.512±0.001 | 2.2085 | 13 |
| | $\Xi^+ + \Xi^-$ | 85.77±0.01 | 0.169±0.01 | 7.89±0.01 | 0.252±0.001 | 4.2645 | 7 |
| 7 | $K^0_S$ | 90.01±0.01 | 0.219±0.01 | 6.41±0.01 | 134.699±0.001 | 16.9862 | 21 |
| | $\Lambda$ | 91.75±0.01 | 0.171±0.01 | 8.25±0.01 | 70.999±0.001 | 19.7664 | 21 |
| | $\Xi^-$ | 92.11±0.01 | 0.179±0.01 | 8.01±0.01 | 7.999±0.001 | 15.8712 | 19 |

*Table 2.* *The values of free parameters, normalization constant and $\chi^2$ for different strange hadrons at different collision energies obtained through the Hagedorn model.*

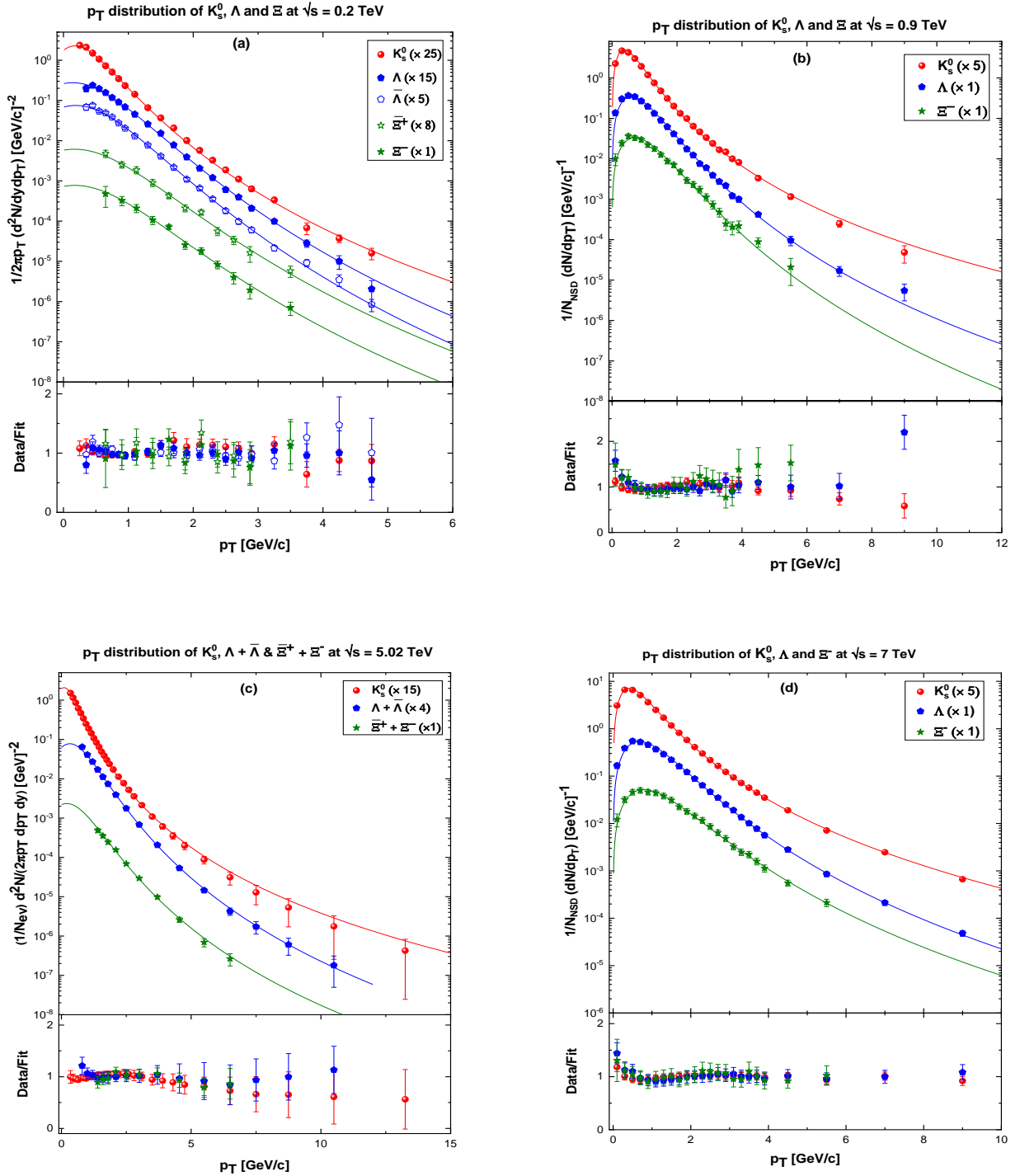

*Fig. 3.* $p_T$ *distribution of strange Hadrons at the different center of mass energies in pp collision, the symbols are used for data points and solid lines represent the fits of Hagedorn model, the lower panel associated with each plot shows the data by fit ratio.*

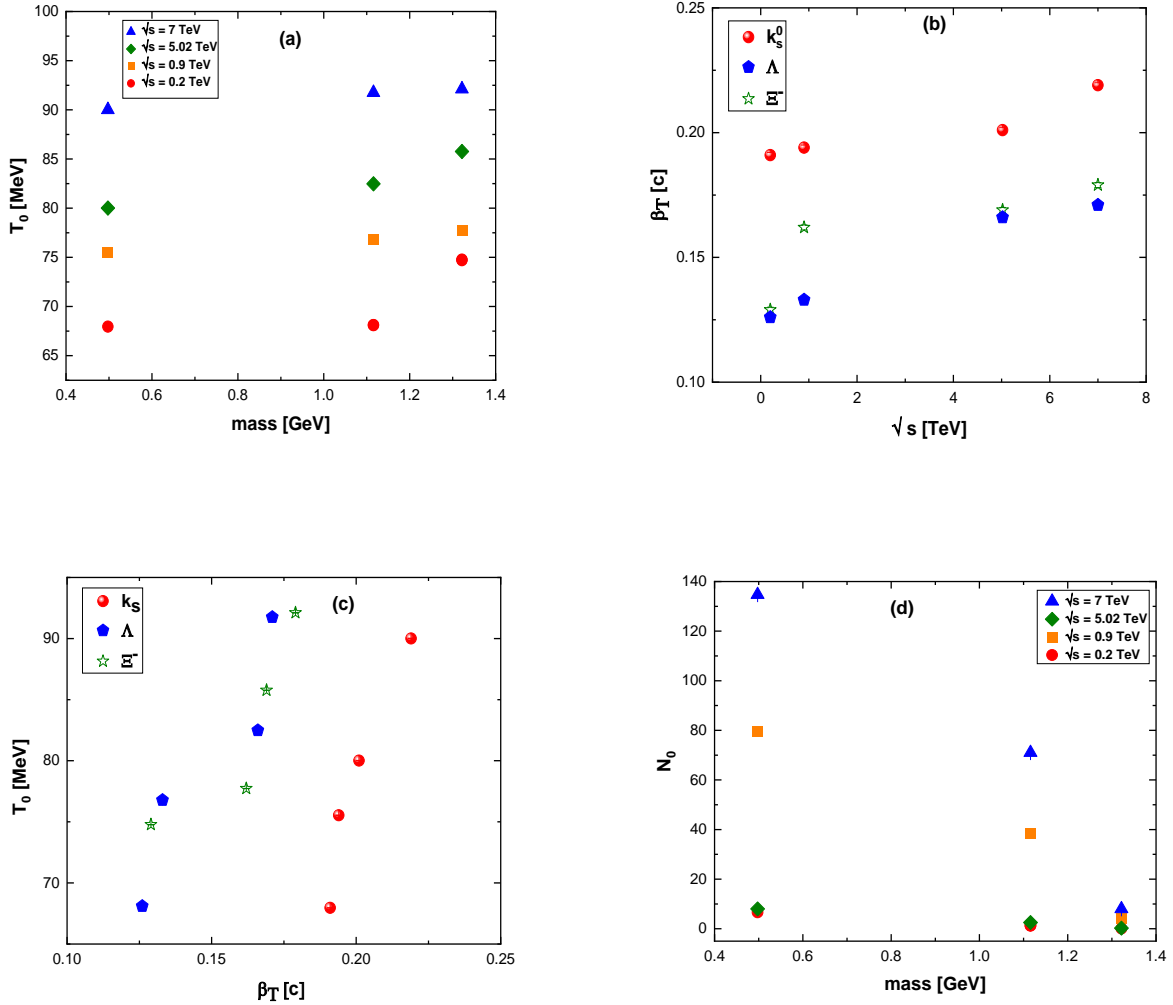

*Fig.4* *Dependence of (a) $T_0$ $m_0$, (b) $\beta_T$ on $\sqrt{s}$ (c) $T_0$ on $\beta_T$ and (d) $N_o$ on $m_0$ at different collision energies using modified Hagedorn model.*

## Conclusion

This research work provides insight into the dynamics of different strange particles produced in pp collisions at different colliding energies. The parameters extracted from the models and different correlations among these parameters can be used in understanding the true nature of the matter under extreme conditions. We are presenting the conclusion of our research in the following points.

(a) In this article, we report analyses of the $p_T$ distribution of strange hadrons by using the Tsallis and modified Hagedorn model with embedded transverse flow velocity for the experimental data of $K_s^0$, $\Lambda$, $\bar{\Lambda}$, $\Xi^+$, $\Xi^-$ at $\sqrt{s} = 0.2$ TeV, recorded by STAR at RHIC, and at $\sqrt{s} = 0.9$ TeV, 5.02 TeV and 7 TeV, recorded by CMS at LHC, in pp collision in rapidity range from 0 to 2. Both models satisfy very well with the experimental data. However, the Tsallis model is found to be a little bit more effective than the Hagedorn model in fitting the experimental data of $p_T$ distribution.

(b) The effective temperature, in the case of the Tsallis model, and kinetic freeze-out temperature, in the case of the modified Hagedorn model, are found to vary linearly with collision energy i.e., smaller temperatures are at $\sqrt{s} = 0.2\ TeV$ and greater are at $\sqrt{s} = 7\ TeV$ because of the greater energy transfer in the high energetic collisions.

(c) These two kinds of temperatures have linear relationship with the particles masses i.e., the greater the mass of the observed particle, the greater is the effective and freeze-out temperatures which shows multi-freeze-out scenario, where each specie of particles decouple from the fireball at different temperatures or at different times. The heavier particles decouple from the system or fireball earlier due to their greater masses or inertia. Because of the greater inertia, the heavier particles are left behind from the fireball while the lighter particles, due to their smaller inertia, are capable to move on with the system during expansion and decouple at later stages.

(d) Inverse correlation between q and m, and T and q have been observed which mean that smaller particles will decouple from the fireball later and hence will attain equilibrium slowly because they have smaller T and larger q compared to heavier particles. The key point behind attaining equilibrium is the value of q. If $q = 1$ the system is said to be in full equilibrium. For heavier particles we have greater temperatures and smaller values of q, close to unity, therefore they attain equilibrium quickly compared to lighter particles.

(e) The multiplicity parameter ($N_0$) is found to decrease with the increase in the $m_0$ which confirms the mass differential freeze-out scenario

(f) A positive correlation between $\beta_T$ and $T_0$ has been observed representing greater $\beta_T$ for those particles having greater $T_0$ and vice versa. At higher energies, the colliding particles transfer greater energy into the system resulting in a higher degree of excitation (greater temperature) of the system, also at the same higher energies a very sharp squeeze occurs in the collision zone or system which results in the rapid expansion (greater $\beta_T$) of the system.